\documentclass[a4paper,fleqn]{cas-dc}
\usepackage[authoryear]{natbib}

\begin{document}
\let\WriteBookmarks\relax
\def\floatpagepagefraction{1}
\def\textpagefraction{.001}

\shorttitle{Early discovery of comets with LSST}

\shortauthors{Inno et~al.}

\title [mode = title]{How much earlier would LSST have discovered currently known long-period comets?}                      



%
\author[1,2]{Laura Inno}[type=editor,
                        orcid=0000-0001-7511-2910]

\cormark[1]

\fnmark[1]

\ead{laura.inno@uniparthenope.it}


\credit{Conceptualization of this study, Methodology, Investigation, Software, Writing- Original draft preparation}

\affiliation[1]{organization={Dep. Science and Technology,Parthenope University of Naples},
    addressline={CDN-IC4, Naples}, 
    city={Naples},
    postcode={I-80143}, 
    country={Italy}}

\author[1]{Margherita Scuderi}
\credit{Data curation, Writing: Original draft preparation}

\author[1]{Ivano Bertini}
\credit{Conceptualization, Methodology, validation}

\author[3]{Marco Fulle}
\credit{Methodology, Validation, Supervision, Writing: Review \& Editing}

\author[4]{Elena Mazzotta Epifani}
\credit{Validation, Supervision, Writing: Review \& Editing
}

\author[2]{Vincenzo Della Corte}
\credit{Validation, Supervision,Writing:Review \& Editing
}

\author[2]{Alice Maria Piccirillo}
\credit{Writing: Review \& Editing}

\author[1]{Antonio Vanzanella}
\credit{Writing: Review \& Editing}

\author[5]{Pedro Lacerda}
\credit{Discussion, Writing: Review \& Editing}

\author[6]{Chiara Grappasonni}
\credit{Resources, Management}

\author[6]{Eleonora Ammanito}
\credit{Resources, Management}

\author[6]{Giuseppe Sindoni}
\credit{Resources, Management}

\author[1]{Alessandra Rotundi}
\credit{Writing: Review \& Editing, Supervision}

\affiliation[2]{organization={INAF-Osservatorio Astronomico di Capodimente},
    addressline={Salita Moraliello, 16}, 
    city={Naples},
    postcode={I-80131}, 
    country={Italy}}

\affiliation[3]{organization={INAF-Osservatorio Astronomico di Trieste},
    addressline={Via Tiepolo 11}, 
    city={Trieste},
    postcode={I-34143 }, 
    country={Italy}}
 
 \affiliation[4]{organization={INAF-Osservatorio Astronomico di Roma},
    addressline={Via Frascati 33, Monte Porzio Catone}, 
    city={Rome},
    postcode={I-00040}, 
    country={Italy}}
    
    \affiliation[5]{organization={Instituto de Astrofísica e Ciências do Espaço, Universidade de Coimbra},
    city={Coimbra},
    country={Portugal}}
    
    \affiliation[6]{organization={Italian Space Agency},
    addressline={via del Politecnico snc}, 
    city={Rome},
    postcode={I-00133}, 
    country={Italy}}

\cortext[Laura Inno]{Corresponding author}



\begin{abstract}
Among solar system objects, comets coming from the Oort Cloud are an elusive population, intrinsically rare and difficult to detect. Nonetheless, as the more pristine objects we can observe, they encapsulate critical cues on the formation of planetary systems and are the focus of many scientific investigations and science missions. 
The Legacy Survey of Space and Time (LSST), which will start to operate from the Vera C. Rubin Observatory in 2025, is expected to dramatically improve our detection ability of these comets by performing regular monitoring of the Southern sky deep down to magnitude 24.5 with excellent astrometry. 
However, making straightforward predictions on future LSST detection rates is challenging due to our biased knowledge of the underlying population. This is because identifications to date have been conducted by various surveys or individual observers, often without detailed information on their respective selection functions. Recent efforts ~\citep[see e.g.][]{vokrouhlicky2019origin} to predict incoming flux of Long Period Comets still suffer of the lack of systematic, well-characterized, homogeneous cometary surveys.
Here, we adopt a different point of view by asking how much earlier~on known comets on long-period or hyperbolic orbits would have been discovered by a LSST-like survey if it was already in place 10 years prior to their perihelion epoch. 
In this case, we are not simulating a real flux of incoming comet, as all comets in our sample reach the perihelion simultaneously, but we can analyze the impact of a LSST-like survey on individual objects. We find that LSST would have found about 40\% of comets in our sample at least 5 years prior to their perihelion epoch, and at double (at least) the distance at which they were actually discovered. Based on this approach, we find that LSST has the potentiality to at least twofold the current discovery rate of long-period and hyperbolic comets.

\end{abstract}




\maketitle

\section{Introduction}
Comets on long-period or hyperbolic orbits approaching the inner Solar System for the first time are the only observational probes of the Oort Cloud, thus containing imprints of the original environment in which the Sun formed. As such, they provide a fundamental linkage between present-time observations and theories of planetary formation. 
Indeed, the dynamical evolution of protoplanetary disks, influenced by the giant planets, helped scatter planetesimals into distant orbits, forming a storage region, the Oort Cloud, where these comets spend the majority of their life until they have a chance to travel towards the inner solar system and eventually become visible to us. Starting from Oort's pioneering work ~\citep{oort1979origin}, our understanding of long period comets has significantly evolved thanks to development of our observing capability and numerical models, from the point of view of the orbital dynamics ~\citep[see e.g.][]{everhart1973examination,francis2005demographics, rickman2010cometary, fouchard2017distribution}, the physical properties~\citep[see e.g.][]{bauer2017debiasing, meech2017}, and the origin and evolution ~\citep[see e.g.][]{brasser2012origin, kaib2011sedna}.
These studies collectively enhanced our comprehension of these object whilst also highlighting the need for further research to address the remaining uncertainties in their population and evolutionary processes.

According to the NASA/Horizon database, there are currently 1133.
\footnote{As of October 2024.} known long-period (LP) or hyperbolic (Hy) comets that have been discovered by a variety of surveys and observers, as shown in Figure~\ref{fig:1}. This highly inhomogeneous catalog does not allow to perform solid statistical population studies, as the selection function of the identified objects remains unknown. This limits all efforts to make accurate quantitative prediction of the flow of such incoming objects.
Recently, \citet{vokrouhlicky2019origin} developed an evolutionary model of the long-period comet (LPC) population based on the observational data collected by \citet[][and references therein]{krolikowska2017oort}  whose result are still limited by the lack of an homogeneous, well-characterized survey to discover and characterize comets (see e.g. the Section 2).
The new Legacy Survey of Space and Time ~\citep[LSST; ][]{2019ApJ...873..111I} of the Vera C. Rubin Observatory is going to change this paradigm, as it will be the first systematic discovery survey of Solar System bodies coming with a well defined selection function. 

After achieving first light next year, LSST will start surveying the entire Southern sky every three nights for ten years in multiple ($u$,$g$,$r$,$i$,$z$,$y$) bands, producing an unprecedented catalog of small bodies that will tenfold or even hundredfold the number of objects currently known for each family ~\citep[see e.g.][etc.]{2009EM&P..105..101J, 2009arXiv0912.0201L, 2010Icar..205..605S, 2015MNRAS.446.2059S, 2016AJ....152..103S, 2016AJ....151..172G, 2017AJ....154...13V, 2018Icar..303..181J, 2018arXiv180201783S, 2019ApJ...873..111I, 2020Icar..33813517F}. 
That's possible because, with respect to precursor surveys carried out at any wave band, LSST features a unique combination of flux sensitivity (down to $r$=24.5 mag in the $r$-band), collecting area (9.6 square degree field of view) 
, and temporal sampling rate (one snapshot every three nights). These properties make LSST the best equipped survey for early detection of potential targets for the ESA mission Comet Interceptor, which will visit for the first time a dynamically new or an interstellar comet ~\citep[][ see also \url{https://www.cosmos.esa.int/web/comet-interceptor/home}]{ci_mission,snodgrass2019}.
The spacecraft will be launched in 2029 and be delivered to the Sun-Earth Lagrange Point L2, where it will reside until a suitable target is identified, and then travel to intercept its trajectory and perform a close flyby, which will happen at a relative velocity between 10 and 70 km/s ~\citep{ci_mission,dellacorte2023}. Given how challenging the mission strategy is, a timely identification of potential targets and an accurate characterization of their dust environment are critical to correctly plan the spacecraft trajectory and ensure its safety. 
As mentioned above, LSST is expected to find thousands of new comets -including at least one interstellar object- per year, and therefore provide an extensive catalogs of potential Comet Interceptor targets~\citep{2023ApJS..266...22S,ci_mission}.
However, providing quantitative estimates of possible discoveries for comets from the Oort-cloud might be difficult, given that the underlying population is not fully understood yet.
As we wait for more robust simulations and/or LSST first data, we can exploit available information of already known long-period comets to estimate how many and how early in-bound they would have been seen by a LSST-kind of survey, defined as a survey done from the same location of the Vera C. Rubin Observatory in a similar way to LSST, if it was operating the ten year prior to their perihelion passage, i.e. with the last year (10$^{th}$) of this LSST-like ideal survey corresponding to the year of the perihelion epoch. 
The main goal of this paper is to provide an answer to this question together with a quick review of the properties of the discovery space for currently known comets. 
A short description of the available sample and its selections is given in section~\ref{s:1}, while the study of their potential observability is described in sections~\ref{s:2}. The limitation of the approach used and the final results from our experiment are discussed in section~\ref{s:3} and ~\ref{s:4}, respectively.

\section{Long-period and hyperbolic comets in the JPL Horizon database}\label{s:1}
We start our analysis by downloading the data available on the Jet Propulsion Laboratory (JPL) Horizon website \footnote{\url{https://ssd.jpl.nasa.gov/tools/sbdb_query.html}} for all objects that belong to the "comets orbit classes" (Hyperbolic Comet, Encke-type Comet,Parabolic Comet, Chiron-type Comet, Jupiter-family Comet,Halley-type Comet, others). Among the initial list of 3,977 objects, 670 have a period equal or longer than 200 years, considered as Long-period (LP) comets (Cs), while 492 are on a hyperbolic orbit with an undefined period (Hy Cs). 
Statistical properties of this sample have already been discussed in several papers, e.g.\cite{yang2022,fernandez2012}.

\begin{figure*}[!t]
\begin{center}
\includegraphics[width=\textwidth]{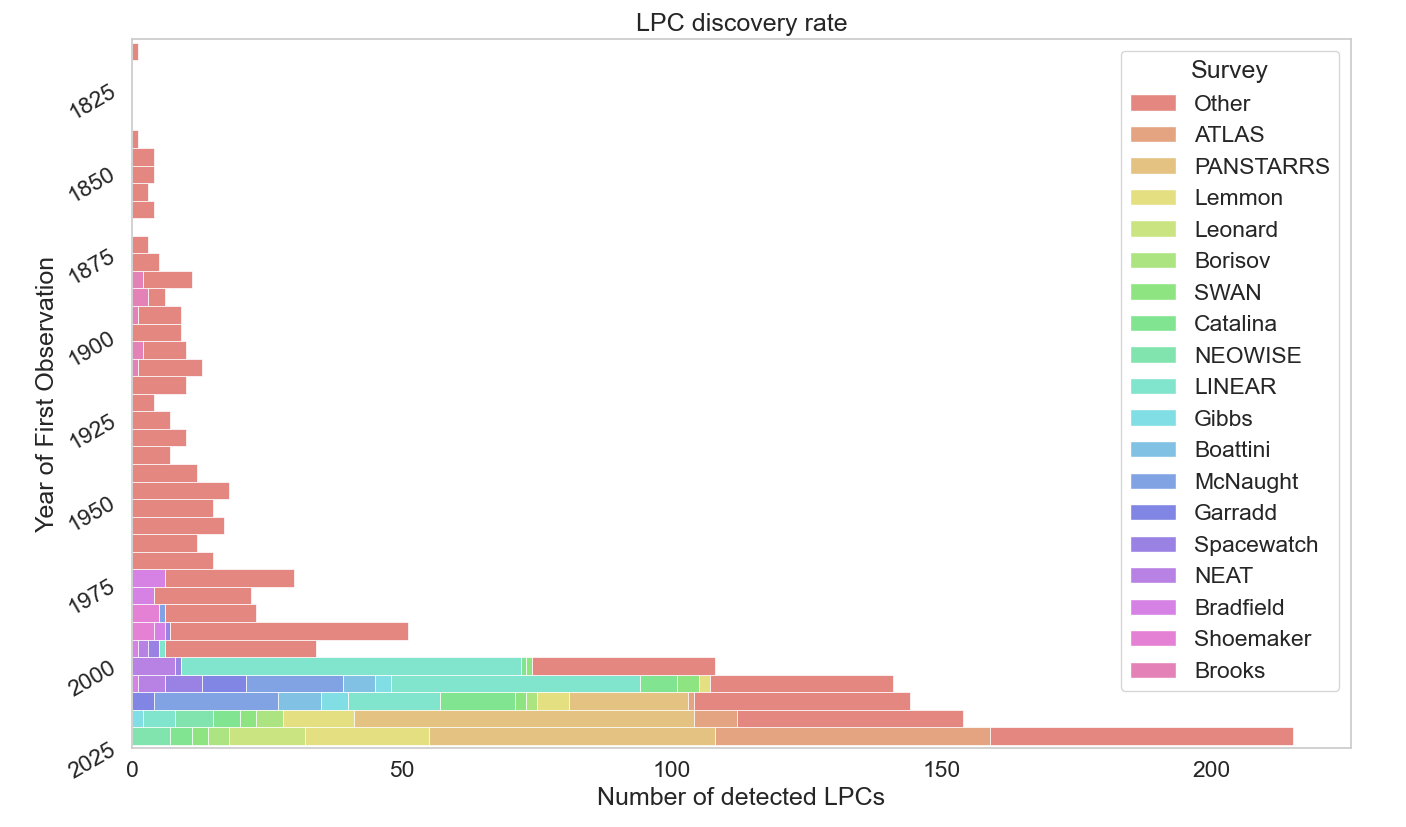}
\end{center}
\caption{Histogram of known long-period (LP) and hyperbolic (Hy) comets in the Horizon database as a function of the year of discovery. Each bin corresponds to five years. Colors correspond to surveys or observatories that have contribute with more than 10 discoveries, as labeled in the legend. We can observe a steady increment of discovery rate over the last two decades, with a current value around 27 LPCs/Hyp Comets per year.}\label{fig:1}
\end{figure*}

\begin{figure*}[!ht]
\begin{center}
\includegraphics[width=\textwidth]{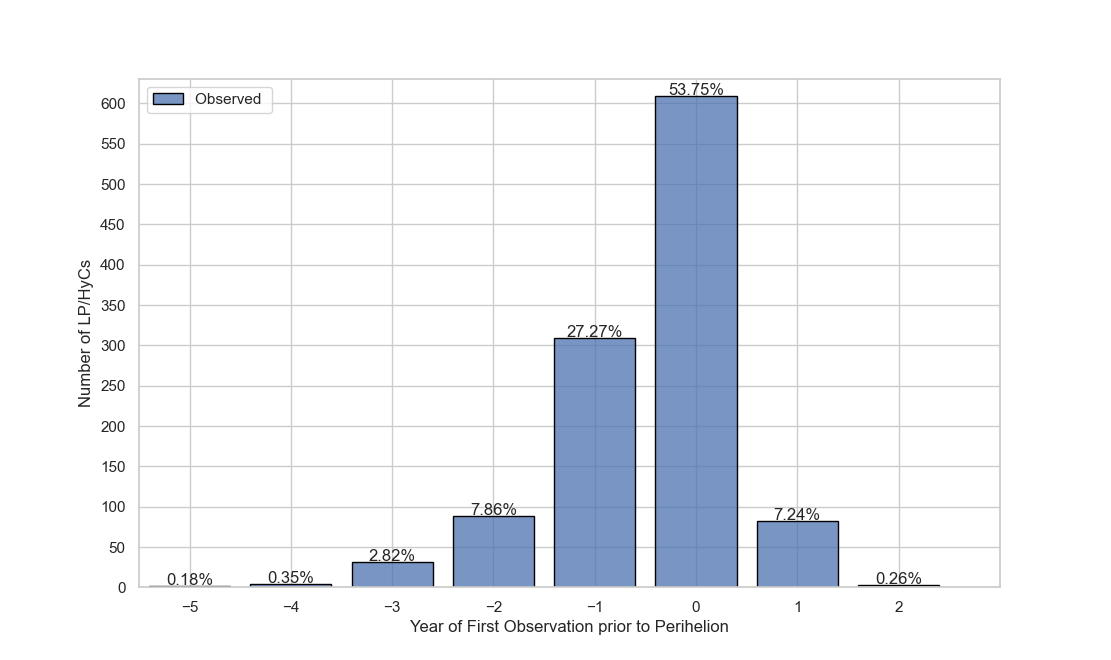}
\end{center}
\caption{Difference in years between the time of first observations and of perihelion. More than half of the LP/Hy comets have been detected only one year prior to perihelion. Moreover, except for the notable case of comet C/2014 UN271 (Bernardinelli-Bernstein) - which has not been included in the plot to improve clarity-, no such comets have been discovered earlier than 5 years with respect to the time of closest proximity to the Earth. Furthermore, there are 72 comets that have been discovered on their outbound orbit, within one year after perihelion.}\label{fig:2}
\end{figure*}

As we are interested in the discovery metric of these objects, we show in Figure~\ref{fig:1} the distribution of the sample in function of the year of their first observation, binned on intervals of five years.
The histogram shows that in the past 20 years, the discovery rate of LPCs has increased over time almost steadily, especially since the 2000, when all-sky surveys such as NEAT, LINEAR, PanSTARRs etc. started to become operative. 
For the same comets we also show the time difference - in years- between their first observation and time of perihelion in Figure~\ref{fig:2}. Except for the notable case of comet C/2014 UN271 (Bernardinelli-Bernstein), which is the farthest comet ever discovered at the heliocentric distance of $r_h \approx 29$ au, 92\% of comets in the sample have been discovered within 4 years prior to their perihelion epoch, with the vast majority ($\approx$85\%) discovered only one year before or after the epoch of closest proximity to us. In fact, 72 comets in the sample have been discovered already on their outbound orbit.

In this work, we want to determine how earlier the same comets would have been discovered if LSST was already operating in the 10 years prior to their perihelion epoch. 
In order to answer this question, we need to know the orbit of the comets and their apparent luminosity as they approached the inner Solar System. 
To this purpose, we use the information accessible through the JPL database, which includes ephemerides and luminosity parameters for 83\% of the initial sample. 
Moreover, to be intercepted by a space mission similar to Comet Interceptor, the target should cross the ecliptic near the Earth, or, more precisely, within a distance that the spacecraft could cover from the point of the L2 orbit from which it will depart. 
In this work, we will consider a potential Comet Interceptor predecessor target as an LP or Hy comet with a $a)$ perihelion distance below 1.2 au and $b)$  a node in the ecliptic plane that could have been reached by the Comet Interceptor spacecraft if it was already flying. 
As we detail in the next paragraph and summarise in Figure~\ref{fig:3}, these conditions restricts the catalog to 211 and 146 comets, respectively. 

\subsection{Comets with perihelion distances closer than 1.2 au }\label{s:2}
Potential targets of Comet Interceptor will need to cross the ecliptic near the Earth, which means that they need to have a perihelion distance $q\leq$ 1.2 au and a small elevation from the ecliptic plane when they approach the Earth, i.e. when they reach heliocentric distances between 0.9 and 1.2 au. The exact conditions for the approach would depend on the relative position of the spacecraft in the L2 orbit and the target at the time of the ecliptic crossing, and on the spacecraft deltav budget, but the requirements above can be used for a general discussion.

\subsection{Final sample composition}
In order to make predictions of the comets' visibility at any given distance, we need to infer their apparent luminosity, by using, as a first order approximation, the following equation: 
\begin{equation}
T= M1 + 5\log \delta + k1\log r_h,
\label{eq:1}
\end{equation}
where $T$ is the total apparent luminosity, $M1$ is the absolute luminosity, $\delta$ is the distance of the object relative to the observer, $r_h$ is the heliocentric distance and $k1$ is the comet total magnitude slope parameter.
The parameters $M1$ and $k1$ are empirically obtained by fitting the observed light curve of that specific comet at some specific time. 
They  are retrieved from JPL Horizons, which fits them using photometric data reported to the Minor Planet Center (MPC). These data include observations from various sources, such as professional and amateur astronomers, and typically span a range of heliocentric distances during the comet's observable period. All details are available directly on the MPC website.
Equation \ref{eq:1} is known as the standard IAU model to describe the comet light curve, a common approach in cometary magnitude modeling, as described in several handbooks on comet observation and orbit modeling, in many papers~\citep[e.g.][]{whipple78,francis2005demographics}, and also adopted by the Minor Planet Center\footnote{\url{https://www.minorplanetcenter.net/iau/mpc.html}} (MPC) and JPL. However, this remains an approximation, since it assumes uniform activity over distance, spherical symmetry for the coma and neglects phase angle effects. We discuss further the shortcoming of this approximation in the next Sections.
However, Equation~\ref{eq:1}, even if simplistic, is a good starting point for bulk analysis of comets, if we keep in mind that for several comets  additional terms accounting for the brightness dependence on the phase angle, and the variation of dust/gas driven activity are needed to fit the observed light-curves. In Horizon, these comets are flagged with a comment in the ephemerides files, and they account for less than 15\% of the entire sample.
The values of $M1$ and $k1$, published on the Minor Planet Center website, and stored in the JPL-Horizons' database, might not be available for all comets and have different degrees of accuracy, depending on the quality and number of the available photometric measurements. 
Hence, the condition "$M1$ ISDEFINED" further reduces our sample from all 146 comets with $q\leq$ 1.2 au to 98. We refer to this sample of comets as "CI's potential Target Predecessors" or CITPs hereinafter. 
Note that in general these are not real potential targets for the space mission, but simply all the objects, discovered from decades ago to year 2024, who share some orbital characteristics with the real, future target. 

\begin{figure*}[h!]
\begin{center}
\includegraphics[width=\textwidth]{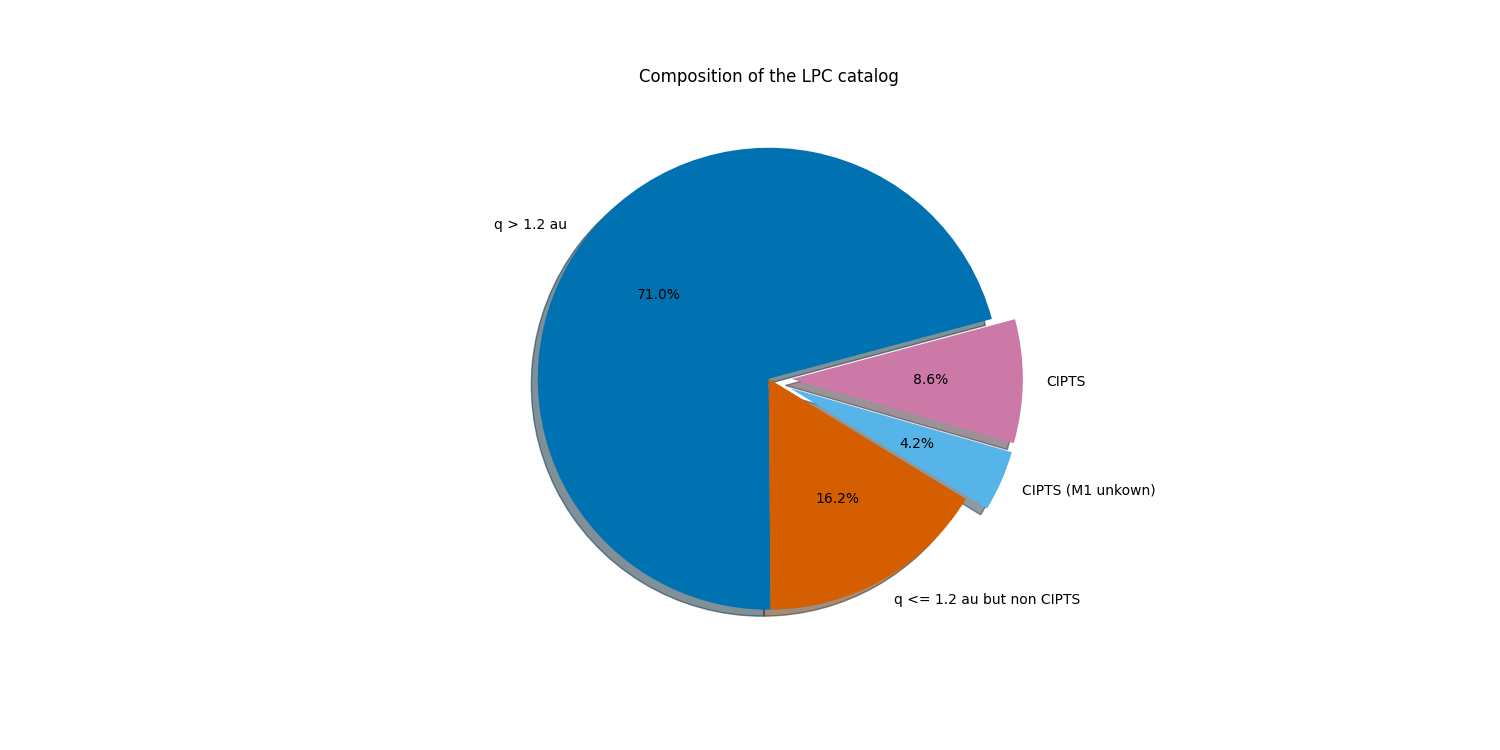}
\end{center}
\caption{Composition of the LP/Hy comet sample used in this work. Among the total 1133 comets, only 329 have a perihelion distance $q$ equal or closer than 1.2 au, and among them only 146 crossed the ecliptic plane at a relative low ecliptic latitude (within 10$^{\circ}$), with 98 of them having known luminosities in JPL: these comets are indicated as CIPTs.}\label{fig:3}
\end{figure*}

However, among them there are six of what are called "virtual targets"\footnote{https://www.cosmos.esa.int/web/comet-interceptor/news}, namely targets just discovered that could have been actually chosen by the mission if it was already in space in August 2022.
The composition of the sample is detailed in the diagram of Figure~\ref{fig:3}, which shows that only about 8\% of the samples has all the dynamical characteristics that we used to define the CIPTs.

\section{Observability of long-period and hyperbolic comets in the ten years before perihelion}\label{s:3}

Once we have the list of comets, we use the python API built-in astropy ~\citep{astropy} to query interactively the Horizon database and obtain ephemerides for each object in our list.  
The Vera C. Rubin Observatory has been assigned an official code from the Minor Planet Center for the coordinates of its observing site ("X05") that can be used to compute the ephemerides.
For each object, we download the ephemerides over a time range of 10 years, considering an interval of 9 years prior to the year in which they reached the perihelion, with a one month step.

The ephemerides also include the total apparent magnitude of the comet $T$ as defined in Equation \ref{eq:1}, converted into the LSST $r$-band magnitude by adding a zero point of $-0.17$~mag ~\citep{willmer2018} . We selected the $r$-band for our analysis because comets are expected to appear brighter in this filter compared to others. Nevertheless, LSST's strategy of alternating between different filters ensures sufficient coverage across all bands. Even if a comet is not initially observed in this, the LSST pipeline combines data from all available filters to identify moving objects, minimizing the impact of band-specific cadence on detection. In our study, we analyze detection timelines using monthly timesteps and focus on yearly discoveries. Consequently, short-term variations, such as whether the initial detections occur in the r-band or another filter, or loss due to bad weather, do not affect our findings.

While it is true that comets are often brighter in the r-band compared to other filters, LSST's strategy of alternating filters ensures that sufficient observations are obtained across multiple bands. Even if an initial observation does not occur in the r-band, the LSST pipeline combines data from all available filters to identify moving objects. This mitigates the impact of band-specific cadence on detection. In our study, we analyzed detection timelines using monthly timesteps and focused on yearly discovery rates. As such, short-term variations in detection timelines, such as whether the initial detections occur in the r-band or another band, do not significantly affect our findings. The robustness of our results stems from the fact that LSST’s observing cadence ensures regular revisit times and sufficient observational coverage in all bands over longer timescales.

\begin{figure*}[!ht]
\begin{center}
\includegraphics[width=\textwidth]{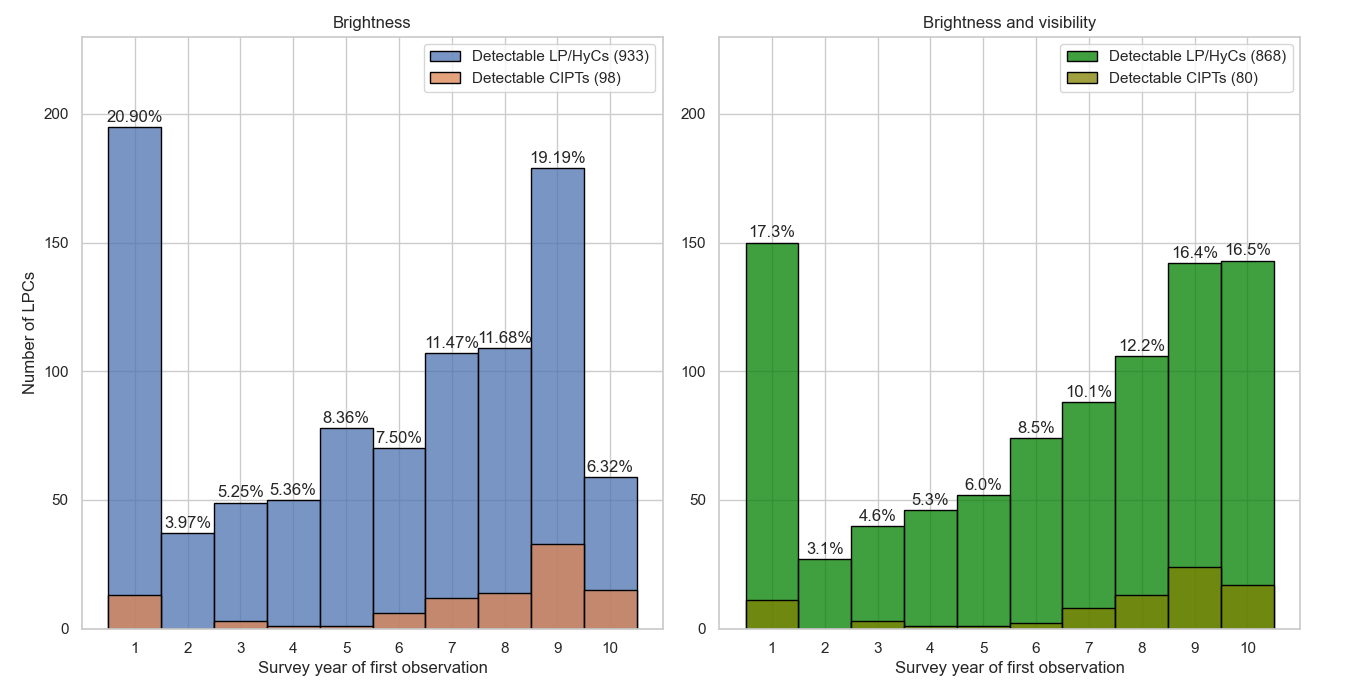}
\end{center}
\caption{Left panel: number of known LP/Hy comets that would have been discovered by a LSST-like survey as a function of survey year, with the 10th year corresponding to the year of perihelion, by taking into account only the brightness limits. The number of CIPTs is also indicated. This figure shows that about 21\% (or 196) of all LPCs would have been already observed in the first year of the survey, and only $\sim$6\% (or 59) in the perihelion year, with an increasing discovery rate in between, as the comets in the sample becomes brighter, with almost half of all comets found already in the first five years.
Right panel: same as left, but adding observability constraints from Cherro Pachòn. When taking into account  the objects' position in the sky, the detection efficiency is reduced, but almost 65\% of all comets are still found at least two years before perihelion. 
}\label{fig:4}
\end{figure*}

If we consider the magnitude range in which LSST will observe, namely between a $r$-band magnitude from 24.5 (nominal single-epoch detection limit) to 14 (single-epoch saturation limit)~\citep[see ][and https://www.lsst.org/scientists/keynumbers]{sciencebook} we find that 82\% of the comets in our sample would have been detected by LSST, the majority of which at least four years before perihelion. Among the observable targets, there are also 67\% of the CI's target predecessors, one fourth of which (i.e. 24) would have been detected at least 5 years before perihelion. The number of LP/Hy Cs discovered as a function of survey year is shown in the left panel of Figure ~\ref{fig:4}.
In this figure we can identify two peaks, one at the first year of the survey and the second one on the ninth, just before perihelion. 
The peak at year one is due to the fact that all comets already bright enough to be detected, would all be found within the first year of observations.
As more and more comets approach the Sun, the number of detections steadily increases, leading to another peak corresponding to the fainter comets just overcoming the luminosity detection threshold.  

Note that in the LSST handbooks there is only one value for the 5-sigma limiting magnitude in the r-band for both point-sources and unresolved sources, so we took that at face value. The impact of neglecting the effect of the extendedness of the source depends both on the actual photometric performance of LSST and on the model adopted to predict the behavior of comets at large heliocentric distances. Indeed, if the activity onset happens at heliocentric distances closer than 6 au, then the extendedness of the coma will have a much lower impact on LSST detection capabilities. 
Recent models  predict  the onset of the activity at distances as large as 100 ~\citep{bouziani2022} or 85 ~\citep{ciarniello23,fulle2023}. In this case the coma will already be spread over hundreds of pixels by the time LSST will detect the comet (if we consider an average traveling time of 1 m/s) and we can suppose that the detection will be delayed. It is not possible with the current knowledge of the LSST system to predict the exact impact of this effect on the detection capability. However, we find that, by using a shallower threshold of 24 mag in the r-band, about one third of the comets in our sample would have been detected one year later than the one reported in Figure ~\ref{fig:4}.

\subsection{Constraints on visibility}
Besides the comet luminosities, we should consider their observability from Cerro Pachòn, by considering that only comets visible at an airmass $\leqslant2.5$ will be detected. 
If we now consider visibility, the number of total comets observed is reduced to 868, but the trend on the early detection is similar: $\approx$40\% of comets would have been observed at least five years before perihelion by LSST, and  84\% of them  at least one year before the perihelion epoch. Among these, 77 are CI target predecessors. The right panel of Figure ~\ref{fig:4} shows the number of LP/Hy Cs detected per survey year, when considering both the limiting magnitude of LSST and their observability from Cerro Pachòn.

\subsection{Constraints on apparent speed}

This means that the detection capability depends on the cadence of the observations, as explained in detail by \citet{2023ApJS..266...22S}. In particular, in Figure 25 of their paper, they show the opposition on-sky motion observed on Earth as a function of heliocentric distances for objects that can be detected with different time gaps between subsequent visits. For two visits separated by 15 minutes the velocity threshold would be 2.0 arc seconds per hour, whilst with a gap of 33 minutes even slower moving objects can be detected by the automated LSST built-in pipeline. 

Detection of moving objects in LSST is done by identifying all sources in the difference image analysis that show an appreciable displacement in two subsequent visits. This means that the detection capability depends on the cadence of the observations, as explained in detail by \citet{2023ApJS..266...22S}. They examined various LSST cadence simulations produced after the implementation of the Feature-Based Scheduler system ~\citep{naghib2019}, starting with version 1.5 in May 2020 and culminating in versions 2.1 and 2.2 in 2022. These simulations explored a wide range of survey strategies, including variations in the time gap between visits and their impact on the detection of moving objects.
In particular, in Figure 25 of  \citet{2023ApJS..266...22S}, they show the opposition on-sky motion observed on Earth as a function of heliocentric distances for objects detectable with different time gaps between visits. For two visits separated by 15 minutes, the velocity threshold would be 2.0 arcseconds per hour, whereas a gap of 33 minutes allows the detection of even slower-moving objects by the automated LSST built-in pipeline. 

By requiring that the on-sky motion is faster than 2.0 arc seconds per hour at the time of the first detection, the number of total objects identified over the 10 year does not change. There is, however, a delay from one month to several years in the epoch of first observation for four comets: C/2024 G7 (ATLAS), C/2013 US10 (Catilina), C/2021 S1 (ISON)and C/2015 V2 (Johnson).

\begin{figure*}[!ht]
\begin{center}
\includegraphics[width=\textwidth]{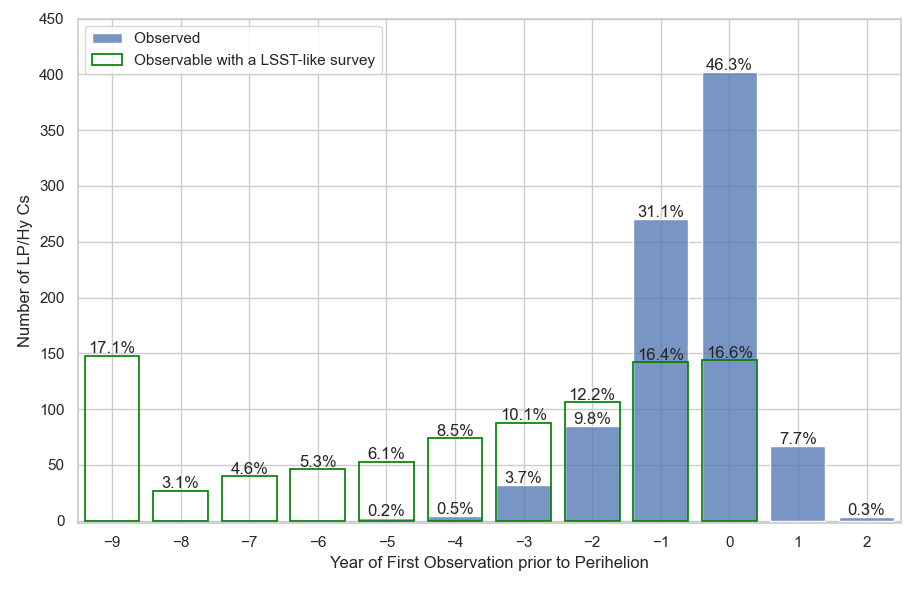}
\end{center}
\caption{Comparison between the histogram of Figure~\ref{fig:2} and the histogram for the same objects discovered from a LSST-like survey when considering all constraints. Note that we simulated only the 10 years prior to perihelion, so outbound discoveries are not included.}\label{fig:5}
\end{figure*}

The final number of LP/Hy comets discovered by taking into account all the above constrains as a function of years prior to perihelion is shown in Figure ~\ref{fig:5}, where we compare this hypothetical sample to the previous histogram in Figure~\ref{fig:2}. Basically, over 45\% of the sample would be discovered at least five year before perihelion, whilst less than 1\% was actually discovered so early on. 

\begin{figure*}[!ht]
\begin{center}
\includegraphics[width=\textwidth]{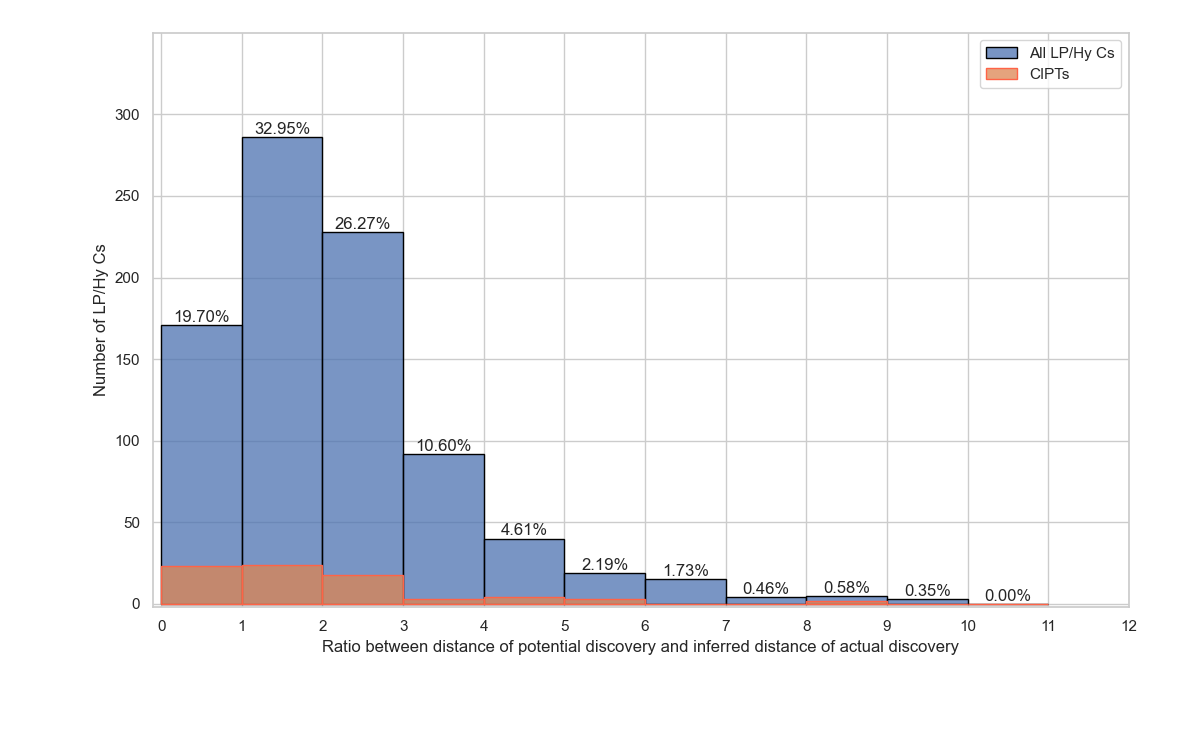}
\end{center}
\caption{Distribution of ratio between the inferred distances at the time of the discovery from a LSST-like survey, $r_{hyp}$, with respect to the inferred distances of LP/Hy comets $r_{fob}$  at the time of the actual first observation, as reported in the Horizon database. 
This means that almost 90\% of the comets would have been observed at farther distances than our current discovery capabilities.  
}
\label{fig:6}
\end{figure*}

By identifying comets earlier on, LSST will be able to detect them at significantly larger distance with respect to the ones at which they have been found up to date. In Figure~\ref{fig:6} we show the distribution of ratio between the distance at which the comet was at the time of first observations according to the ephemeris computed by the Horizon software,$r_{fobs}$, and the distance at which they would have been found from a LSST-like survey $r_{hyp}$. 

The majority of comets will be observed systematically farther, and 46\% of them will be detected at double the distance, at least. 
This means that the discovery space of LSST will be open to much further/fainter objects, that are currently passing undetected. 

However, we find that 144 comets, including 17 CIPTS, would still be observed for the first time in the year of perihelion even for a LSST-like survey and basically at similar distance at which they were indeed discovered. This happens because the magnitude predicted backwards for these objects might be much fainter than the actual one at the discovery. 
In general, one needs to be careful when using Equation \ref{eq:1}, to predict the expected luminosity of comets, especially if applied at times that are far from the original observations. 

\subsection{Comparison between predictions and observations}

\begin{figure*}[!ht]
\begin{center}
\includegraphics[width=\textwidth]{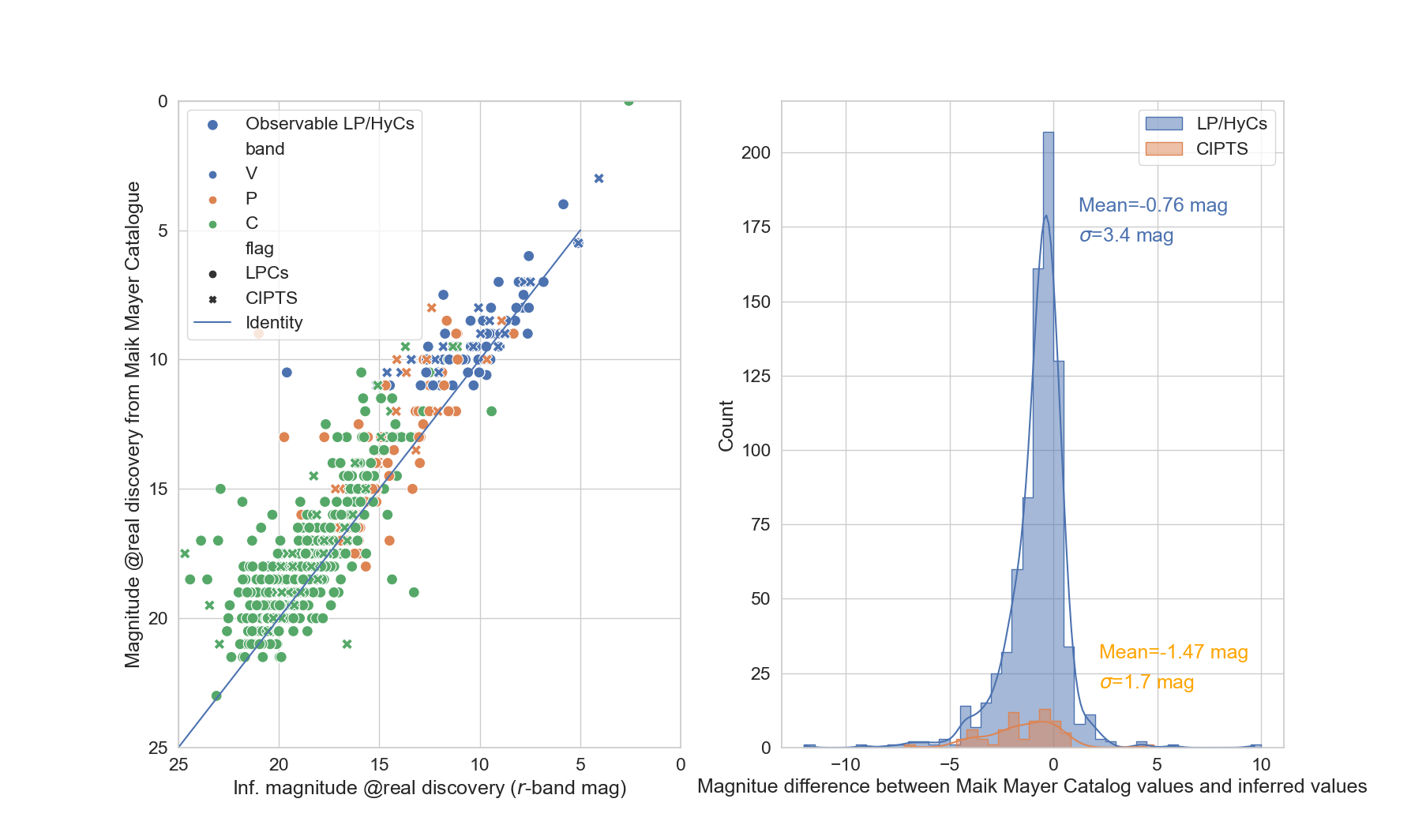}
\end{center}
\caption{Left: Visual magnitude listed in the Maik Mayer Catalogue of Comets as a function of the one inferred from Equation~\ref{eq:1} at the time of discovery. As shown in this plot, inferred magnitude are systematically fainter. Indeed, the difference between these two values can be up to 12 magnitudes, with a mean around -1 magnitude and a standard deviation larger than 3 magnitude, as shown in the histogram in the right panel. 
}\label{fig:7}
\end{figure*}

In this work, we used the JPL-Horizon calculation to infer the luminosity of the comets at earlier time with respect to the actual observations. Therefore, we can compare their expected luminosity with the measured one at the time of discovery. Unfortunately, an homogeneous, standard catalog that includes this information is not available given that comets are usually discovered by different surveys and/or amateur astronomers with non-standard exposure times and settings. 
We tried to address this issue by using the Meyer Catalogue of comet discovery (Maik Meyer, personal communication\footnote{\url{https://www.comethunter.de/project.html}})
to perform a general comparison between the inferred luminosity of the comets in our catalog and the one listed there, shown in the left panel of Figure~\ref{fig:7}.
For about 80\% of the comets in common, the Meyer Catalogue lists a brightness measured with CCD detectors. However, even in these cases the measured value of the brightness might be affected by several effects, i.e. shortness of exposure, different filters etc. When comparing the inferred magnitude at discovery with the one in the Mayer Catalogue, we found that the latter are systematically brighter of one magnitude on average, which is within the error budget but still indicates that we are systematically underestimating the luminosity of our targets and hence the true capability of an "ideal" LSST-like survey to detect objects earlier on.
Moreover, 10\% of the total sample and 20\% of the CIPT sample show differences larger than 3 magnitudes. 
This mostly reflects the limitations of the reliability of available photometric measurements to make systematic analysis of the sources for statistical purposes, such as the prediction of incoming flow of potential targets for the Comet Interceptor space missions, an obstacle that LSST will help to overcome.
In the next section, we discuss the comparison between the available data and the predictions from Horizon for the CI virtual target and other relevant cases.

An alternative method to Equation\ref{eq:1} to model the comet luminosity is based on the so-called $Af\rho$ parameter, introduced by \citet{ahearn84}, and defined as: 
\begin{equation}
Af\rho= \frac{4r_h^2 \Delta^2 10^{0.4(m_S-m_C)} }{\rho}\  [\rm{meters}];,
\label{eq:2}
\end{equation}
where where $A$ is the average grain albedo, $f$ the filling factor in the aperture field of view, $\rho$ the linear radius of the aperture at the comet, expressed in meters, $r_h$ the heliocentric distance expressed in au, $\Delta$ the geocentric distance in meters, $m_s$ the Sun magnitude, and $m_c$ the cometary magnitude, measured in the aperture of radius $\rho$.
This quantity is widely used to estimate the sunlight reflected by dust in the comet’s coma and is proportional to the dust loss rate ~\citep[see][and reference therein]{weiler2003dust,woodward2015sofia,epifani2016photometry,fulle2022comets}. In cases where the cometary coma is in a steady state, the value for $Af\rho$ becomes independent of the aperture size, making it particularly useful for consistent measurements of the cometary dust continuum. 
If we have a reliable set of $Af\rho$, and we know how it evolves with $r_h$ for each specific comet, we can invert Equation~\ref{eq:2} to obtain an estimate of the comet magnitude, for a given choice of $\rho$.
Unfortunately, comprehensive studies of $Af\rho$ values for long-period comets are limited, which forced us to adopt Equation~\ref{eq:1} to predict the comets' point-source equivalent magnitude with a discrepancy that can be up to $\approx$3 magnitudes, as shown in Figure~\ref{fig:7} ~\citep[see also][]{fernandez1999inner}.
In \cite{fulle2022comets}, homogeneous measurements of $Af\rho$ for five LP comets are presented. 
We used the values in their Table 5 in order to estimate the luminosity of comet  C/2017 K2 (PANSTARRS) and C/2020 F2 (ATLAS) at their nominal discovery distance, 16 au and 9.82 au, respectively. 
The observations reported in the Minor Planet Center database for these comets indicate a magnitude of 19.7 and of 19 at the time of discoveries. 
By using the $Af\rho$ values at $R_h$=10 au of 55$\pm$10 and 30$\pm$5, and assuming $\rho$=2.5 x 10$^7$ m, we find a predicted total luminosity in the $r$-band of 20.8 mag (at $\Delta$=$2.4 10^{12}$m and $r_h$= 16 au) and 18.12 mag (at $\Delta$=$1.5 10^{12}$m and $r_h$= 10 au), which are both consistent with the observations.  
For those comets, the Horizon predictions also show a good agreement with the reported observations, by indicating 19.9 mag and 18.7 mag, respectively. 
Whilst this agreement is promising, only two examples do not provide any statistics, but we plan to collect a larger, homogeneously measured sample of comets and address this comparison between light-curve modeling and $Af\rho$ based magnitudes in a forthcoming paper.

\section{Limitations and Results}\label{s:4}
This study has severe limitations, the strongest of which is that our dataset is not representative of the flux of incoming comets from the Oort Cloud, and cannot be used to make predictions on the absolute number of future discoveries from LSST. 
Moreover, the brightness assumed for the LP/Hy Cs as a function of the distance by using Equation\ref{eq:1} relies on strong assumptions and has three major sources of uncertainty: the limited accuracy with which $M1$ and $k1$ are obtained from photometric observations, the fact that these values may change over time even for the same object (Lacerda et al. in preparation), and as the drivers of  cometary activity change as a function of heliocentric distance ~\citep[see e.g.][]{feaga2007,fougere2016,fulle2022comets}, the luminosity of individual comets can also depends on their initial chemical composition, the sizes of active areas, specific volatile species, and their depths within the cometary nucleus ~\citep{meech2017}.

We also used the current orbit parameters to compute the ephemerides in the ten year before perihelion, but these parameters also suffer from the uncertainty to which their are determined, related to the limited number of observations and/or their temporal arc. More importantly, we are also neglecting any temporal evolution produced by gravitational and non-gravitational effects that might perturb the orbit as the comet travel within the inner Solar System.

Hence, with such limitations, the reader could wonder what is the relevance of the work presented here. Ideally, we would like to determine the flux of incoming Long-Period comets in order to make quantitative predictions on LSST performance and, in turn, on the number of CIPTs. In principle, this can be achieved by linking together prediction from a dynamical model, and the survey selection function. Several attempts have been made in order to predict the incoming flux of comets, such as, \cite{everhart1967comet,francis2005demographics,sosa2011updated,vokrouhlicky2019origin}. 
If we restrict our sample to include only comets with $q\leq4$, absolute magnitude $H\leq$10.9 mag and have been discovered from 2020 to date, we find about 41 objects, which is roughly consistent with the prediction from \citet{francis2005demographics}, assuming a surveyed area that includes all sky and a detection efficiency of 100\%, which is not realistic.
On the other hand, if we rely on theoretical models, determination of the incoming flux from numerical simulations are size-limited, and we should assume a size-magnitude relation in order to compute the detection efficiency, a quite complex task, as discussed in detail by \citet{vokrouhlicky2019origin}.
Given the relevance of the topic, many efforts are underway to predict the impact of LSST's observing strategy and capabilities on synthetic populations of solar system bodies. In particular, within the LSST Solar System Science Collaboration, the development of the survey simulator enabled the release of LSST Data Preview 0.3, which includes simulated solar system catalogs after one and five years of survey~\citep[see \url{https://dp0-3.lsst.io}and mentions in][]{2023ApJS..266...22S}. 

In this work, we do not predict future performance but instead analyze the potential impact of LSST on the detection of already known objects. The main findings from our analysis are as follows:
\begin{itemize}
    \item Only ten percent of the entire sample would have been missed by a Southern Hemisphere survey;
    \item None of the comets would have been missed due to the Wide-Fast-Deep standard cadence, although some discovery times might have been delayed by up to four years;
    \item Forty percent of comets, including nearly all CI virtual targets, would have been detected at distances at least double those of their actual discovery.
\end{itemize}

\begin{table*}[!ht]
    \centering
    \begin{tabular}{lcccccccc}
    \toprule
    Name & $m_r$ & $\Delta_{hcr}$[au] & h$_{hcr}$ & dyear$_{b.p.}$ & dyear$_{LSST}$ & a$_{LSST}$ & rh$_{LSST}$ &rh$_{real}$ \\
    \midrule
    C/2021 F1 Lemmon-PANSTARRS  &           14.80 &                    2.08 &                         -4.46 &                     -1 &         9 &   1.74      & 4.25 &          5.72 \\
    C/2023 P1 Nishimura  &           13.08 &                    1.65 &                         -5.12 &                      -1 &        9 &         2.05 &          7.7 & 4.23 \\
    C/2022 E3  ZTF  &            8.36 &                    0.30 &                          7.17 &                     -1 &         1 &         2.09 &         22.23 & 4.87 \\
    C/2023 R2 PANSTARRS  &            13.85 &                    1.85 &                         -2.76 &                     -1 &         7 &         1.77 &          10.28 & 6.9\\
    C/2023 H2 Lemmon  &     12.22      &    0.20                &  6.57                   &    0                  &     9     &    1.41      &   6.28     & 4.22   \\
    C/2023 V5 Leonard &      18.58      &         0.23            &       -3.74                   &     0                 &    10     &     1.14     &   4.76    & 1.15  \\
    \bottomrule
    \end{tabular}
    \caption{Table of Comet Interceptor virtual targets. $m_r$= magnitude at first crossing of the ecliptic plane, $\delta_{hcr}$= delta at first crossing of the ecliptic plane, h$_{hcr}$= ecliptic latitude at first crossing of the ecliptic plane, dyear$_{b.p.}$= discovery year before perihelion, dyear$_{LSST}$= survey year of potential LSST discovery, a$_{LSST}$= airmass at LSST discovery, rh$_{LSST}$= distance at LSST discovery in au,rh$_{real}$= distance at real discovery in au obtained on the basis of our model.}
    \label{tab:CI_targets}
\end{table*}

\section{Predictions for Comet Interceptor Virtual Targets}\label{s:new}
Predictions based on Equation~\ref{eq:1} are influenced by the limitations discussed in previous sections. However, the accuracy of the model in predicting the luminosity of an individual comet depends significantly on the comet’s unique properties. Here, we present a detailed analysis of specific comets of particular interest, as they are potential targets for the Comet Interceptor (CI) mission.
The list of these comets and the parameters computed for them are given in Table~\ref{tab:CI_targets}.

\subsection{C/2021 F1 (Lemmon-PANSTARRS)} 

Comet C/2021 F1 (Lemmon-PANSTARRS) was discovered on March 19, 2021 by the Pan-STARRS and Mount Lemmon Survey teams. At the time of its discovery, the comet had a magnitude of about 19.7. Its distance from the Sun at discovery was approximately 6.9 au, beyond the orbit of Jupiter.
For this comet, the Horizon calculations provides a $r$-band magnitude of 23.1 mag at a distance of 4.9 au, thus showing a mismatch of $\approx$2 au and $\approx$3 mag between the predictions and the observed data. The photometric data for this comet are available on the MPC database and on the COBS\footnote{\url{https://cobs.si/home/}} database, and they both show a significant scatter of about 2 magnitude for the early observations. Hence, the difference we found is consistent with the propagated uncertainty on the fitted values. By underestimating the luminosity at a given distance, we are delaying the time at which the LSST-like survey would have found the same object, which indeed is just about a few months before the actual discovery.

\subsection{C/2023 P1 (Nishimura)} 

Comet Nishimura (C/2023 P1) was discovered on August 12, 2023 by amateur astronomer Hideo Nishimura.
At the time of discovery, its magnitude was estimated to be around 10.4 and the comet was approximately 1.0 au from the Sun. On the MPC database, a magnitude of  20.6 in the $r$-band was reported on the 10th of March 2023 from the ZTF survey, whilst our model predicts a r-band magnitude of 19 mag for the same date. It also finds a magnitude of 13.4 at distance 1.0 AU for the date of 12th of August. 
Similarly to the case of C/2021 F1, also in this case the photometric observations from the different observatories and observing sites reported on the MPC database show a scatter larger than 2 magnitude, with values as low as 14.4 mag in the $G$ band.
Therefore, we think that the predictions from our basic model are consistent for this comet, within the uncertainty associated to the actual observations. 
In our prediction, a LSST-like survey would have been able to discover this comet one year in advance with respect to its actual discovery, and at distance of about 8 au. However, if our model underestimates the luminosity of similar comets, this hypothetical survey would have been able to find it even earlier on.

\subsection{C/2022 E3 (ZTF)} 
Comet C/2022 E3 (ZTF) was discovered at $\approx$4.3 au. Our backwards predictions, based on Equation~\ref{eq:1}, find that it was at $\approx$4.8 au at the time of discovery, with a predicted magnitude of 17.5 instead of the observed 17.3. In this case, there is a very good agreement between our prediction and the observed data. Our model shows that a LSST-like survey would have discovered the same comet at the much farther distance of $\sim$21 au.

\subsection{C/2023 R2 (PANSTARRS)} 
Comet C/2023 R2 (PANSTARRS) was discovered on September 7, 2023 by the Pan-STARRS survey.  At the time of discovery, its magnitude was approximately 20.1, making it quite faint. The comet was at a distance of $\approx$5.89 au from the Sun when it was discovered, which places it beyond the orbit of Jupiter. Horizon provides a  distance of 4.7 au in September 2023 with an $r$-band magnitude of 20.1. Hence, we find a good agreement between the backward propagation and the real data also in this case. The hypothetical LSST-like survey would have discovered this comet at double the distance, $\sim$10 au.

\subsection{C/2023 H2 (Lemmon)}
Comet C/2023 H2 (Lemmon) was discovered by the Mount Lemmon Survey on April 25, 2023. At the time of its discovery, it was approximately at 2.74 au from the Earth and 3.72 au from the Sun. The comet had an apparent magnitude of about 19 mag at the time of discovery. Horizon predictions find the comet at a distance of 2.9 au from the Sun with an apparent magnitude of 20.1. The LSST-like survey would have then detected the same comet at double the distance, around 6 au.

\subsection{C/2023 V5 (Leonard)}
Comet C/2023 V5 (Leonard) was discovered on November 1, 2023. At discovery, it was located at approximately at the distance from Earth of about 1.25 au and from the Sun of $\sim$1.92 au.The apparent magnitude at discovery was around 18.5. The Horizon predictions provide a good match, with a magnitude of 18.6 at the distance of 1.15 au. For this comet, intrinsically faint and which is moving at high rate (850" per hour), even a survey as powerful as LSST would have been able to anticipate the discovery only of a few months.

\section{Conclusions}\label{s:5}

LSST is poised to revolutionize our understanding of the Solar System, along with other key areas in astrophysics. In this exercise, we aimed to explore the impact that an 'ideal' LSST-like survey would have had on the detection of Oort Cloud comets—especially potential targets for ESA's Comet Interceptor mission—if it had been operational for the ten years leading up to each comet’s perihelion. Our results indicate that over 150 comets could have been detected as early as ten years before their perihelion by this 'ideal' survey, with 87\% of them discovered at distances at least twice as far as their actual discovery distances.

While these findings are not predictive of LSST's future discovery rate, as we did not consider the actual flux of incoming comets, they highlight the survey's impressive discovery potential. To accurately forecast the discovery rate, we will need to determine the number of comets LSST will detect and normalize it by the surveyed area, utilizing a survey simulator.
The simulator has already been developed by the LSST Rubin Solar System Pipelines and Commissioning team and it will be able to provide an accurate selection function for LSST discoveries, once that the data will start to be collected.

Indeed, this is what makes LSST a unique machine to explore the solar system small bodies populations: each data release will include a comprehensive set of sky simulations, such as those from the LSST Operations Simulator, which predict the survey's observational coverage of different sky regions under various conditions. These simulations are essential for developing a detailed and accurate selection function, accounting for both the spatial and temporal coverage of the survey. Furthermore, each data product will include all intermediate-step images used for processing, including calibration files, enabling users to reconstruct each observational detail backward if needed. The first LSST yearly data release, anticipated one year after first light (just recently achieved!), will feature a 'golden catalog' exclusively of LSST-discovered objects, with a well-defined selection function. Thus, one year from now, we will be able to make statistically robust predictions for the survey's following nine years.

We thank; the Italian Space Agency (ASI) within the ASI-INAF agreements I/024/12/0 and 2020-4-HH.0; the PhD program PON “Ricerca e Innovazione” 2014-2020, DM n.1061 (10/08/2021) and n. 1233 (30/07/2020).
We also thank the anonymous referees for their thoughtful review of the manuscript that helped to improve its clarity.
This research has made use of data and/or services provided by the International Astronomical Union's Minor Planet Center.

\printcredits

\bibliographystyle{cas-model2-names}

\bibliography{cas-sc-inno_arx.bib}


\end{document}